\documentclass[letterpaper, 10 pt, conference]{ieeeconf}  

\IEEEoverridecommandlockouts                              
\usepackage{graphics} 
\usepackage{epsfig} 
\usepackage{mathptmx} 
\usepackage{times} 
\usepackage{amsmath} 
\usepackage{amssymb}  
\usepackage{amsfonts, color}
\usepackage{graphicx, subfigure}

\title{\LARGE \bf
Randomly Distributed Delayed Communication and
Coherent Swarm Patterns
}

\author{Brandon Lindley  and Luis Mier-y-Teran-Romero and Ira B. Schwartz 
\thanks{This work was supported by the Office of Naval Research and the
  National institutes of Health.}
\thanks{Brandon Lindley  is an NRC postodctoral fellow at the US Naval
  Research Labooratory, Code 6792, Washington, DC 20375 USA,         {\tt\small brandon.lindley.ctr@nrl.navy.mil}}%
\thanks{L. Mier-y-Teran-Romero is an NIH post doctoral fellow at the Naval
  Research Laboratory. 
{\tt\small luis.miery@nrl.navy.mil}}
\thanks{I. B. Schwartz is at the US Naval
  Research Labooratory, Code 6792, Washington, DC 20375 USA 
{\tt\small Ira.schwartz@nrl.navy.mil}}
}

\begin{document}

\maketitle
\thispagestyle{empty}
\pagestyle{empty}

\begin{abstract}

Previously we showed how delay communication between globally coupled
self-propelled agents causes new spatio-temporal patterns to arise
when the delay coupling is fixed among all agents~\cite{Forgoston08}. In this paper,
we show how discrete, randomly distributed delays affect the dynamical patterns.
In particular, we investigate how the standard deviation of the time delay
distribution affects the stability of the different patterns as well as the
switching probability between coherent states.

\end{abstract}

\section{INTRODUCTION}
Numerous recent investigations have been devoted to the study of interacting
multi-agent or swarming systems in various natural and engineering fields of study. Investigations
of interacting systems have revealed the {emergence of}  highly
complex dynamic behaviors  in space and time which arise even though the dynamics of a single agent is quite simple~\cite{Erdmann05}. In particular, these multi-agent swarms can
self-organize in complicated spatio-temporal patterns that depend on the
details of the inter-agent interactions. These investigations have been
  motivated by and  had an
impact on many diverse biological systems such as bacterial colonies,
schooling fish,
flocking birds, swarming locusts, ants, and pedestrians \cite{Budrene95, Toner95, Parrish99, Topaz04, hebling1995}. In this paper, we are
interested in the application that biological analogies have on the
design of systems of autonomous, inter-communicating robotic systems \cite{Leonard02,Justh04, Morgan05, chuang2007} and mobile sensor networks \cite{lynch2008}.

There is great interest to design agent-interaction
protocols to carry out robotic motion planning, consensus and cooperative
control, and spatio-temporal formation. One methodology is to combine inter-agent potentials with external
ones in order to achieve multi-agent cooperative motion in a manner that is
not too sensitive with respect the number of agents. Some
important applications making use of scalable numbers of agents are: obstacle avoidance \cite{Morgan05}, boundary tracking
\cite{hsieh2005,Triandaf05},  environmental sensing \cite{lynch2008,lu2011},
decentralized  target  tracking \cite{chung2006}, environmental consensus
estimation \cite{lynch2008,Jad2006} and task allocation \cite{mather2011}.

Authors have employed very diverse approaches in the study of multi-agent
systems. Some authors have described the swarms at the individual level,
writing their models in terms of ordinary  differential equations (ODEs) or
delay differential equations (DDEs) to describe their trajectories
\cite{vicsek95,flierl99,Justh04}. The addition of noise on the swarm's dynamics introduces even richer
behavior, such as noise-induced transitions between different coherent
patterns \cite{Erdmann05, Forgoston08}. The study of noisy swarm dynamics has
benefited from tools from statistical physics applied to both first and second
order phase transitions that have been found in the formation of coherent states \cite{aldana07}.

One important aspect of the understanding and design of  space-time behavior
  in communicating robotic systems is that of time delay. Time delay arises in
latent communication between agents, as well as actuation lag times due to inertia.
Time delays can have interesting and surprising dynamical consequences in a system, such as
large-scale synchronization \cite{Englert11, Zuo10, Hache10}, and have been used successfully for control purposes
\cite{Atay07, Konishi10}.  Many of the initial time-delay studies focused on the case of one or a few
discrete time delays. Recently, more complex situations have been considered
such as the case of having several \cite{Ahlborn07} and random time-delays
\cite{Wu09, Marti06}. Another interesting case is that of distributed time
delays, i.e. when the dynamics of the system depends on a continuous interval in
its past instead of on a discrete instant \cite{Omi08}.

In the case of swarming systems in stochastic environments, it has been observed that
the introduction of a discrete communication time delay induces a transition from one
spatio-temporal pattern to another as the time delay passes a certain
threshold \cite{Forgoston08}. It was shown in \cite{Forgoston08} how
the complex interplay  exists between the attractive coupling and the
time delay in the transitions between different spatio-temporal patterns
\cite{MierIROS11,MierTRO11}. Time delays in robotic systems have been also studied in
the contexts of consensus estimation \cite{Jad2006} and task allocation
\cite{mather2011}; in the latter, the time delays originate from the period of
time required to switch between different tasks.

In  this paper, we consider a swarming model with discrete, randomly
distributed time delays.  We
explicitly show how a distribution of delay times perturb the dynamics from
the single discrete case delay case analytically. We illustrate the dynamical
effects of delay distributions with varying width and show that the system is
bistable, and very sensitive to choice of initial starting conditions.

\section{Swarm Model}\label{sec:SM}

We investigate the dynamics of a two-dimensional system of $N$ identical
self-propelling agents that are attracted to each other in a symmetric
manner. We consider the attraction between agents to occur in a
time-delayed fashion, due to the finite communication speeds and
information-processing times. Specifically, we focus on the situation in which the
time-delay is nonuniform across agents: there is one time delay for every
pair of agents $\tau_{ij}(=\tau_{ji})$, for particles $i$ and $j$. The time
delays $\tau_{ij}$'s are time-independent and are drawn independently from a random
distribution $\rho_\tau(\tau)$. The swarm dynamics are  described by the following governing equations:
\begin{subequations}\label{swarm_eq}
\begin{align}
\dot{\mathbf{r}}_i =& \mathbf{v}_i,\label{swarm_eq_a}\\
\dot{\mathbf{v}}_i =& \left(1 - |\mathbf{v}_i|^2\right)\mathbf{v}_i -
\frac{a}{N}\mathop{\sum_{j=1}^N}_{i\neq j}(\mathbf{r}_i(t) -
\mathbf{r}_j(t-\tau_{ij})),\label{swarm_eq_b}
\end{align}
\end{subequations}
for $i =1,2\ldots,N$. The position and velocity of the $i$th agent at time $t$
are denoted
by $\mathbf{r}_i$ and $\mathbf{v}_i$, respectively. Each agent has self-propulsion and frictional drag forces given by the
expression term $\left(1 -  |\mathbf{v}_i|^2\right)\mathbf{v}_i$. The coupling constant $a$ measures the strength of the attraction between
 agents and the communication time delay between particles $i$ and $j$ is
 given by $\tau_{ij}$. Note that in the absence of coupling agents tend to
 move in a straight line with unit speed as time tends to infinity.

\section{Mean Field Approximation}

We carry out a mean field approximation of the swarming system by switching to
particle coordinates relative to the center of mass and disregarding the noise
terms. The center of mass of the swarming system is given by
\begin{align}                                                                                    
\mathbf{R}(t) = \frac{1}{N} \sum_{i=1}^N\mathbf{r}_i(t).                                         
\end{align}
We can decompose the position of each particle into
\begin{align}\label{pos_decomp}                                                                  
\mathbf{r}_i = \mathbf{R} + \delta \mathbf{r}_i,  \qquad i =1,2\ldots,N,                         
\end{align}
where we'll have
\begin{align}\label{linear_dep}                                                                  
\sum_{i=1}^N\delta\mathbf{r}_i(t) = 0.                                                           
\end{align}
 Inserting Eq. \eqref{pos_decomp} into the second order system
 equivalent to Eq. \eqref{swarm_eq} and simplifying we get
\begin{align}\label{CM1}                                                                         
\ddot{\mathbf{R}} + \delta\ddot{\mathbf{r}}_i =& \left(1 -
  |\dot{\mathbf{R}}|^2 -                
2\dot{\mathbf{R}}\cdot \delta\dot{\mathbf{r}}_i -                                                
|\delta\dot{\mathbf{r}}_i|^2\right)(\dot{\mathbf{R}} +                                           
\delta\dot{\mathbf{r}_i})\notag\\                                                                
& - \frac{a(N-1)}{N}\bigg(\mathbf{R}(t) +                                   
\delta\mathbf{r}_i(t)\bigg) \notag\\
&+ \frac{a}{N}\mathop{\sum_{j=1}^N}_{i\neq j}
\left(\mathbf{R}(t-\tau_{ij}) + \delta\mathbf{r}_j(t-\tau_{ij})\right),                             
\end{align}
Summing Eq. \eqref{CM1} over $i$ and using Eq. \eqref{linear_dep}, we get
\begin{align}\label{CM}                                                                          
\ddot{\mathbf{R}}=& \left(1 - |\dot{\mathbf{R}}|^2
  -\frac{1}{N}\sum_{i=1}^N|\delta\dot{\mathbf{r}}_i|^2\right)\dot{\mathbf{R}}
 \notag\\
&-\frac{1}{N}\sum_{i=\
1}^N\left(2\dot{\mathbf{R}}\cdot \delta\dot{\mathbf{r}}_i +                                      
|\delta\dot{\mathbf{r}}_i|^2\right)\delta\dot{\mathbf{r}_i}   \notag\\                           
& -a\frac{N-1}{N}\mathbf{R}(t) + \frac{a}{N^2}\sum_{i=1}^N \mathop{\sum_{j=1}^N}_{i\neq j} \left(\mathbf{R}(t-\tau_{ij}) + \delta\mathbf{r}_j(t-\tau_{ij})\right).                                
\end{align}
We now make some approximations on the terms with the double sums. For the
displacements from the center of mass, we have
\begin{align}\label{dr_approx}
\frac{a}{N^2}&\sum_{i=1}^N \mathop{\sum_{j=1}^N}_{i\neq j}  \delta\mathbf{r}_j(t-\tau_{ij}) = \frac{a(N-1)}{N^2}\sum_{j=1}^N \frac{1}{N-1}                                                      
\mathop{\sum_{i=1}^N}_{i\neq j} \delta\mathbf{r}_j(t-\tau_{ij}) \notag\\ 
&\approx                                                                                
\frac{a(N-1)}{N^2}\int_0^\infty \sum_{j=1}^N                                         
\delta\mathbf{r}_j(t-\tau)\rho_\tau(\tau)d\tau = 0,
\end{align}
since $\sum_{j=1}^N\delta\mathbf{r}_j(t-\tau) =0$ by Eq. \eqref{linear_dep}. In
passing from the discrete to the continuous averaging above, we argue
as follows. The expression $\frac{1}{N-1}\mathop{\sum_{i=1}^N}_{i\neq j} \delta\mathbf{r}_j(t-\tau_{ij})$ is
the average of $\delta\mathbf{r}_j(t)$ at the $N-1$ times $t-\tau_{ij}$. Since
$N\gg 1$ and the times $\tau_{ij}$ are distributed with density
$\rho_\tau(\tau)$, this is approximately equal to $\int_0^\infty \delta\mathbf{r}_j(t-\tau)\rho_\tau(\tau)d\tau$.

Similarly,
\begin{align}\label{R_approx}
\frac{a}{N^2}\sum_{i=1}^N \mathop{\sum_{j=1}^N}_{i\neq j}\mathbf{R}(t-\tau_{ij}) \approx \frac{a(N-1)}{N} \int_0^\infty \mathbf{R}(t-\tau)\rho_\tau(\tau)d\tau.
\end{align}

In a purely heuristic manner, we neglect all fluctuation terms 
$\delta\mathbf{r}_j(t)$ in the dynamics
of the center of mass  and obtain the mean field approximation:
\begin{align}\label{mean_field}                                                                  
\ddot{\mathbf{R}}=& \left(1 - |\dot{\mathbf{R}}|^2 \right)\dot{\mathbf{R}}
-a\left(\mathbf{\
R}(t) - \int_0^\infty \mathbf{R}(t-\tau)\rho_\tau(\tau)d\tau \right).                                                               
\end{align}
where we approximated $\frac{N-1}{N}\approx 1$, since we are considering large
numbers of agents.

\section{Bifurcations in the Mean Field Equation}

The behavior of the system in the mean field approximation in different
regions of parameter space may be better understood by using bifurcation
analysis. This mathematical technique will allow us to show how the parameter
plane of coupling constant $a$ and mean time delay $\mu_\tau$ is divided into regions with different dynamical behaviors.

First we show that Eq. \eqref{mean_field} has a uniformly
translating solution $\mathbf{R}(t) = \mathbf{R}_0 + \mathbf{V}_0 \cdot
t$, where $\mathbf{R}_0$ and $\mathbf{V}_0$ are constant,
two-dimensional vectors. Inserting the uniformly translating state into
Eq. \eqref{mean_field}, we get
\begin{align}\label{unif_trans}                                                        0=& \left(1 - |\mathbf{V}_0|^2 \right)\mathbf{V}_0
-a \int_0^\infty \tau \rho_\tau(\tau)d\tau\mathbf{\
V}_0,                                             
\end{align}
since $\int_0^\infty \rho_\tau(\tau)d\tau=1$. Hence, the speed
$|\mathbf{V}_0|$  of the
uniformly translating state must satisfy
\begin{align}\label{unif_trans}
  |\mathbf{V}_0|^2 = 
1-a\int_0^\infty \tau \rho_\tau(\tau)d\tau = 1 - a\mu_\tau,                                   
\end{align}
where $\mu_\tau$ is the mean of the $\rho_\tau$ distribution. We note that the
direction of motion and starting point $\mathbf{R}_0$ are arbitrary.

The other state of interest is the stationary state $\mathbf{R}(t) =
\mathbf{R}_0$, for an arbitrary constant vector $\mathbf{R}_0$. In the two-parameter space $(a, \mu_\tau)$, the hyperbola $a \mu_\tau = 1$ is in fact a pitchfork bifurcation line on which
the uniformly translating states are born from the stationary state.

The linear stability of the stationary state is determined by the solutions to the characteristic
equation of Eq. \eqref{mean_field}:
\begin{align}\label{char_eq}
\mathcal{D}(\lambda) = a\left(1- \int_0^\infty \rho_\tau(\tau) e^{-\lambda\tau}d\tau\right) - \lambda + \lambda^2 = 0,
\end{align}
and so involves the Laplace transform of the distribution $\rho_\tau$.

In our numerical simulations of system \eqref{swarm_eq}, we considered  a
truncated Gaussian distribution:
\begin{align}\label{rho_tau}
\rho_\tau =
\begin{cases}
{\cal N} e^{\frac{(\tau - \tau_0)^2}{2\tau_1^2}} & \text{if } \tau \ge 0 \\
0 & \text{if } \tau < 0,
\end{cases}
\end{align}
where ${\cal N}$ is the normalization constant. Note that because of the
truncation, $\tau_0$ and $\tau_1$ are only
approximately equal to the mean and standard deviation of $\rho_\tau$ and ${\cal N}$ is only approximately $1/\sqrt{2\pi  \tau_1^2}$.

We approximate the Laplace transform of the truncated Gaussian distribution by
extending the integration range to the whole real line and taking ${\cal N}
\approx \sqrt{2\pi  \tau_1^2}$. In addition, we approximate the mean and
standard deviation of $\rho_\tau$ as $\mu_\tau \approx \tau_0$ and
$\sigma_\tau \approx \tau_1$, respectively. The result is
\begin{align}
\int_0^\infty \rho_\tau(\tau) e^{-\lambda\tau}d\tau \approx e^{\lambda \mu_\tau
  + \lambda^2 \sigma_\tau^2/2}.
\end{align}

\begin{figure}[t!]
\begin{center}
\subfigure{\includegraphics[scale=0.26]{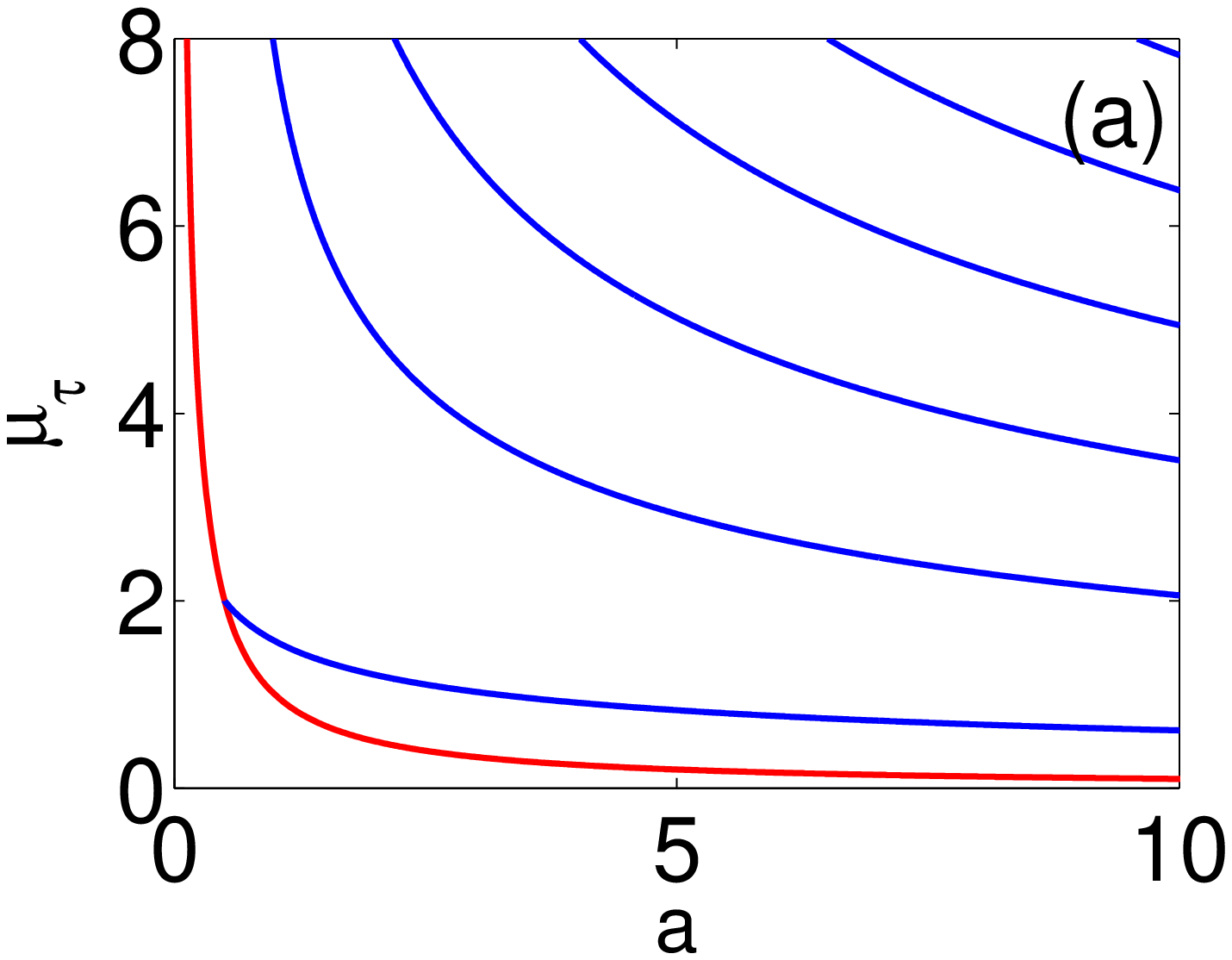} \label{Bifurcations_a}}
\subfigure{\includegraphics[scale=0.26]{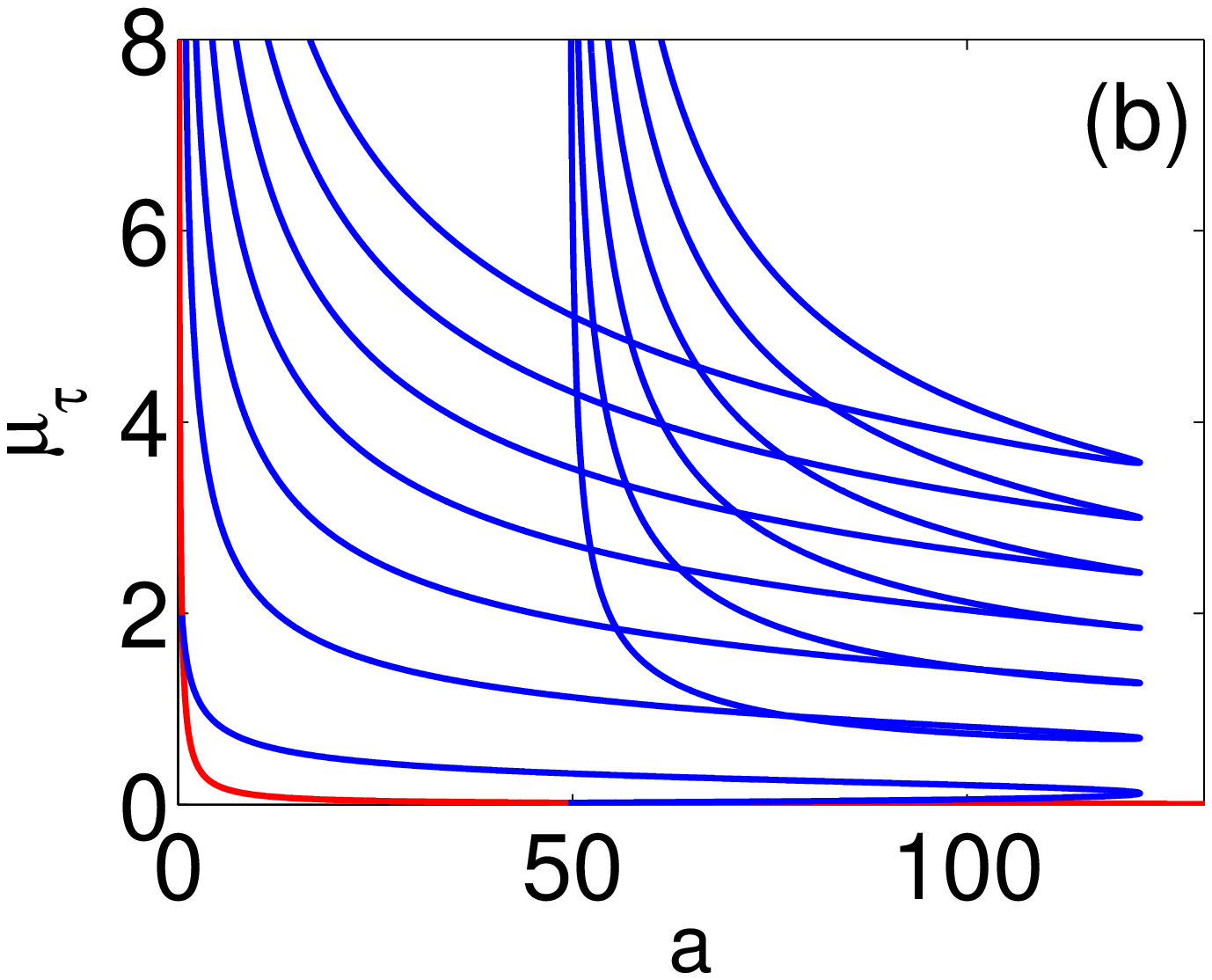} \label{Bifurcations_b}}
\subfigure{\includegraphics[scale=0.26]{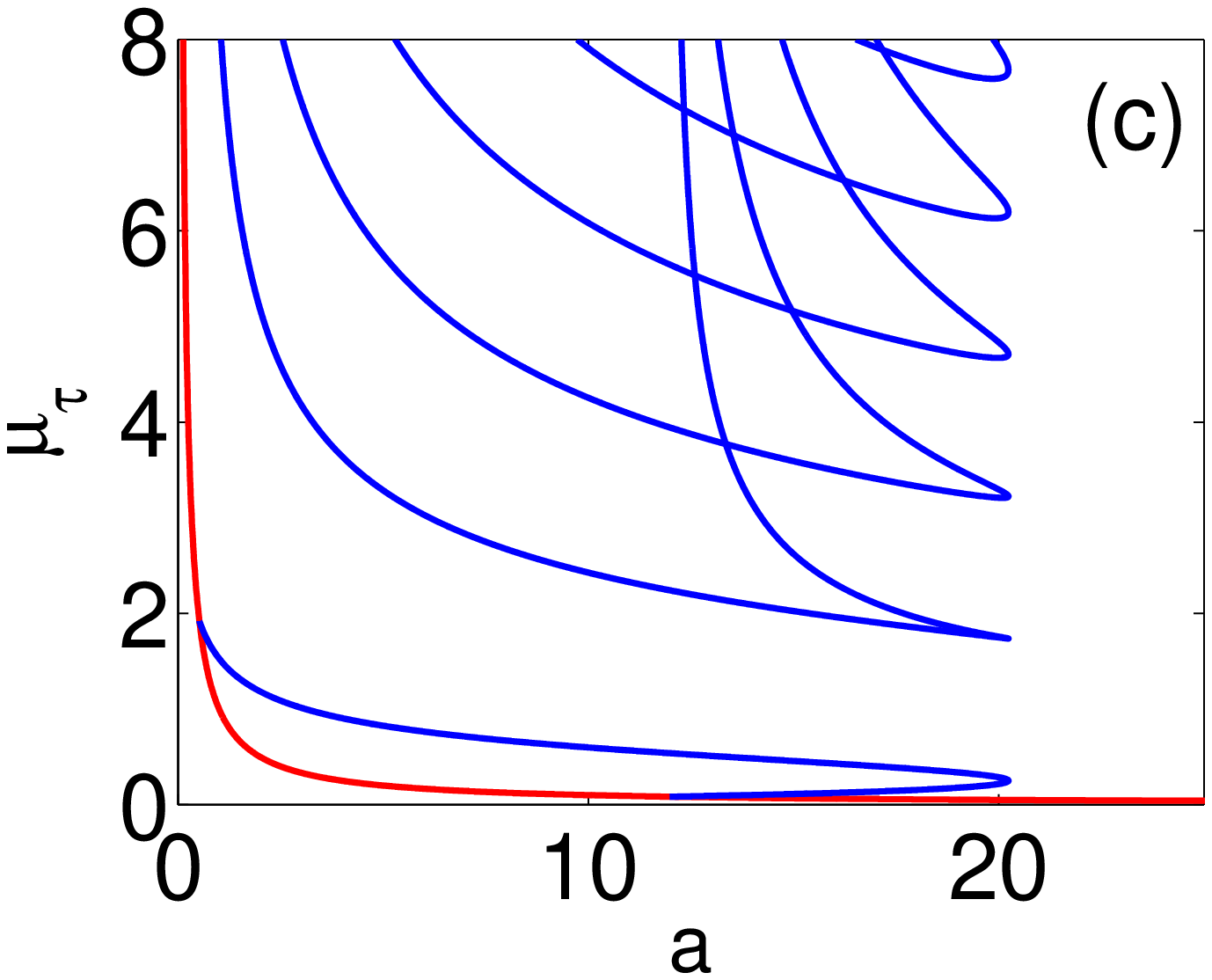} \label{Bifurcations_c}}
\subfigure{\includegraphics[scale=0.26]{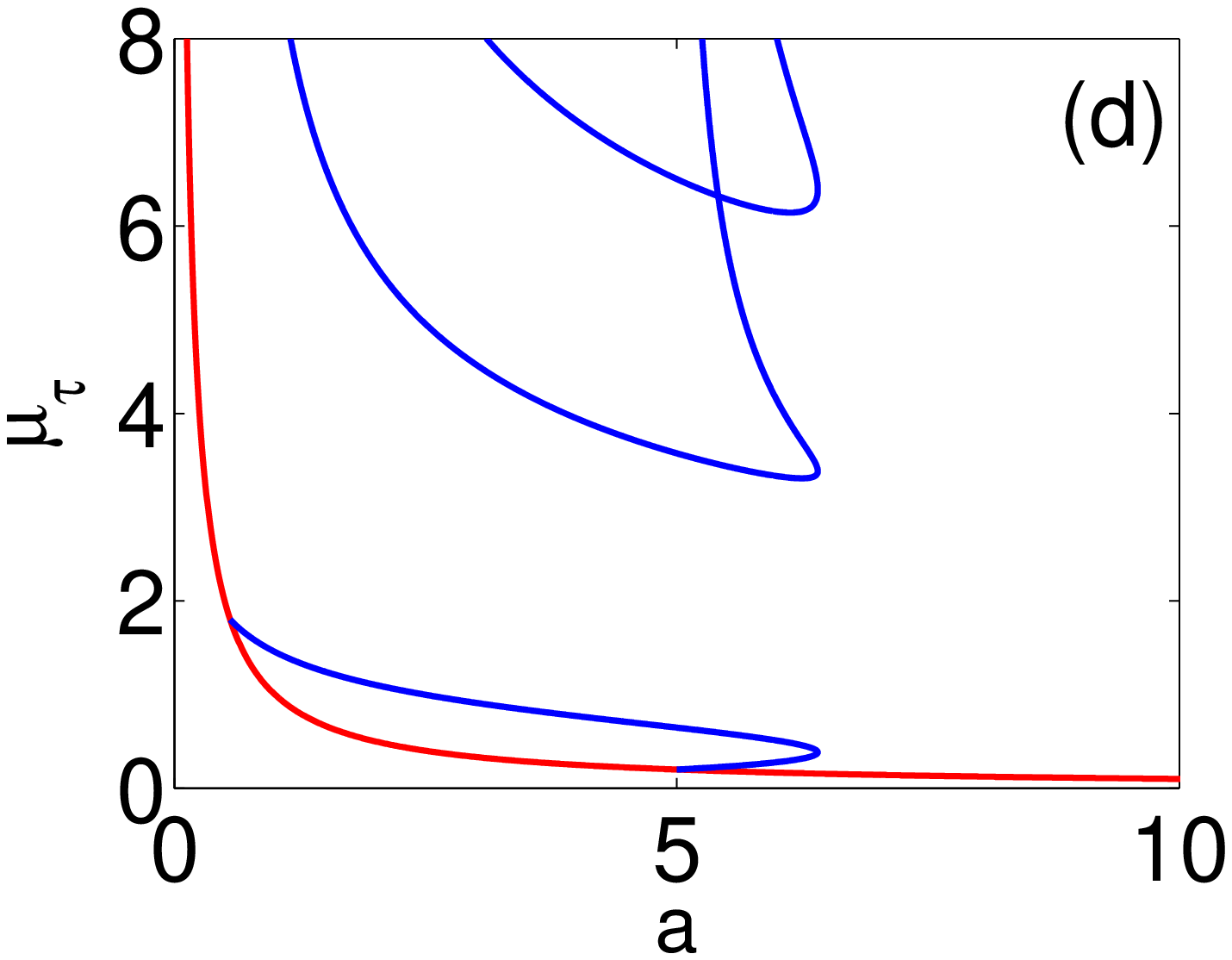} \label{Bifurcations_d}}
\caption{Hopf (blue) and pitchfork (red) branches in $a$ and  $\mu_\tau$
  space. The standard deviations of the time-delay distribution $\rho_\tau$
  for the panels (a) through (d) are 0, 0.2, 0.4 and 0.6,  respectively. Note
  the change of scale in the abscissae.}\label{Bifrucations}
\end{center}
\end{figure}

We use the above approximation to the Laplace transform of $\rho_\tau$ to
search for Hopf bifurcation curves in the $(a, \ \mu_\tau)$ plane, by taking
$\lambda = i\omega$ in the characteristic equation \eqref{char_eq}. The
equation $\mathcal{D}(i\omega)=0$ is equivalent to:
\begin{align}\label{char_eq_hopf}
 a - \omega^2 - i\omega = a e^{i\omega \mu_\tau
  - \omega^2 \sigma_\tau^2/2},
\end{align}
from which we obtain the Hopf curves parameterized by $\omega$:
\begin{subequations}\label{hopf}
\begin{align}
a_H(\omega) =& \frac{\omega\left(\omega \pm \sqrt{e^{-\sigma_\tau^2 \omega^2}(1+\omega^2)
       - 1}\right)}{1 - e^{-\sigma_\tau^2 \omega^2}},\\
\mu_{\tau H}(\omega) =& \frac{1}{\omega}\left(
\arctan\left(\frac{\omega}{a_H(\omega)-\omega^2}\right) + 2n\pi\right), \ n = 0,1,\ldots.
\end{align}
\end{subequations}
In the above expression for $\mu_{\tau H}(\omega)$, the branch of tan in $(0,\pi)$ should be used, since the
complex number on the left hand side of Eq. \eqref{char_eq_hopf} is always on
the top half plane. This family of Hopf curves labeled by $n$, together with the pitchfork bifurcation
curve $a\mu_\tau =1$ are shown in Figure \ref{Bifrucations}, for various
values of $\sigma_\tau$.

When $\sigma_\tau=0$, the system exhibits a {degenerate} point at
$a=1/2$, $\mu_\tau = 2$  (Fig. \ref{Bifurcations_a}), where the Hopf
bifurcation frequency becomes zero. {This is similar, but not equivalent to a
Bogdanov-Takens bifurcation as is known from previous
work \cite{MierIROS11, MierTRO11}. Since the point on the Hopf curve in a
two-parameter bifurcation plane occurs when the Hopf frequency becomes zero,
we define this point as a Zero Frequency Hopf ({ZFH}) point.}

For $\sigma_\tau>0$, this {{ZFH}}  point shifts and a second {ZFH} point appears at
$a\rightarrow \infty$ and $\mu_\tau\rightarrow 0$. The location of the two {ZFH}
points in the $(a, \ \mu_\tau)$ plane is given by $(a_{{ZFH}}^{(\pm)},
1/a_{{ZFH}}^{(\pm)})$, where $a_{{ZFH}}^{(\pm)} = \frac{1}{\sigma_\tau^2}(1 \pm \sqrt{1 - \sigma_\tau^2})$.

When $\sigma_\tau=0$, the behavior of the mean field in the vicinity of the {ZFH}
point is relatively well understood \cite{MierTRO11,MierIROS11} and is as follows
(see Fig. \ref{Bifurcations_a}). In the region between the pitchfork and the first
member of the Hopf family, the stationary state is stable. A simulation of the
full system \eqref{swarm_eq} with parameters in this area reveals that indeed the center of mass of the
agents comes to rest as time progresses and the particles spread themselves
along a ring with radius $1/\sqrt{a}$. Roughly half of the particles move
clockwise and the other half counterclockwise. Along the first Hopf curve, a
stable limit cycle is born and the center of mass begins to oscillate
periodically on a circular orbit. Below the pitchfork bifurcation curve
$a\mu_\tau=1$, the translating state is stable. Finally, we mention that there
is a region of bistability in the parameter region above the {ZFH} point $(1/2, \
2)$ between the pitchfork curve
$a\mu_\tau=1$ and the curve $a\mu_\tau^2=2$ (not shown), where the center of
mass can either translate or rotate. On the curve $a\mu_\tau^2=2$ there is a
global bifurcation where the radius of the orbit diverges and the limit cycle disappears.

The above discussion helps us understand the bifurcation planes in
Figs. \ref{Bifurcations_b} through \ref{Bifurcations_d}. Most significantly,
we see that the parameter region where the stationary state is stable
decreases in size as the width of the time delay distribution widens. Hence
the system has a higher tendency to behave in an oscillatory manner for wider
time delay distributions. This effect has been corroborated in numerical
simulations (results not shown).

\section{Numerical Simulations}

We analyze the dynamics of system \eqref{swarm_eq} by solving the system of
DDEs numerically. We use Heun's method together with quadratic Lagrange
interpolation to evaluate the time-delayed terms of
Eqs. \eqref{swarm_eq}. Overall, the numerical method is second order with
respect to the step-size $\Delta t$. For all simulations we take the agents to
be uniformly distributed in a random fashion within the unit box $0\leq x
\leq 1$ and $0\leq y \leq 1$, and each particle is initially at rest
$\mathbf{v}_j=0$. Moreover, since we are interested in investigating the
time-asymptotic behavior, for all numerical experiments the time of
integration is long enough to allow transients to decay.

In \cite{MierIROS11, MierTRO11} it was shown that for the parameter set $a=2$, $\tau=2$ (fixed
delay) that the system exhibited a bistable set of solutions. In the rotating
state solution, all particles collapse to a point and that cluster of
particles rotates around a fixed center in a circular orbit. The other
possible stable solution is a ring state, in which all particles distribute
themselves uniformly along a circle and orbit around its center at unit speed. Interestingly, not all particles traverse the ring in the same
direction; roughly half move clockwise and half move anti-clockwise. We will
now examine these two states, but with random delays given by the truncated
Gaussian distribution in Eq. \eqref{rho_tau}.

\begin{figure}[t]
\centering
\subfigure[]{\includegraphics*[scale=.26]{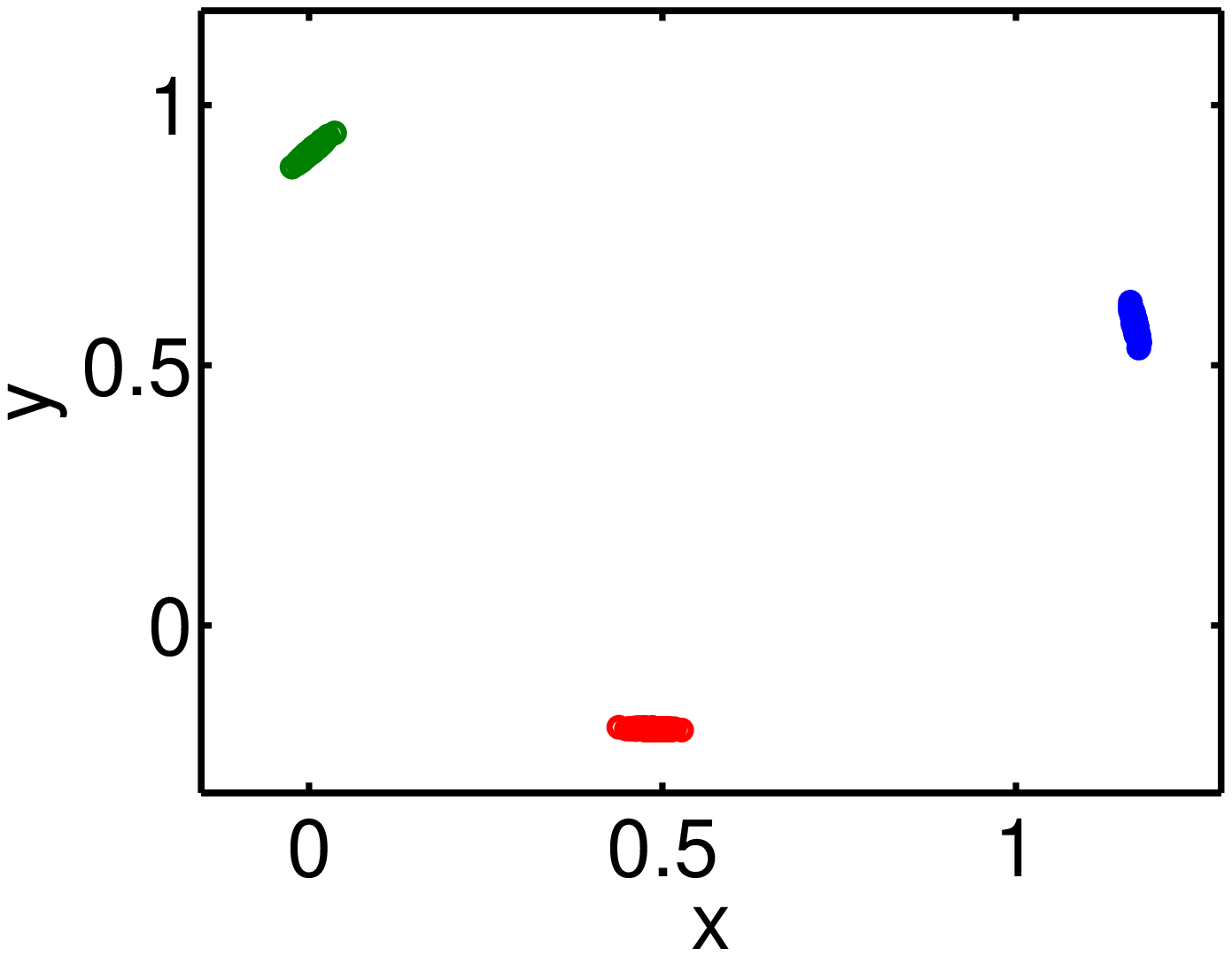}\label{fig1a}}
\subfigure[]{\includegraphics*[scale=.26]{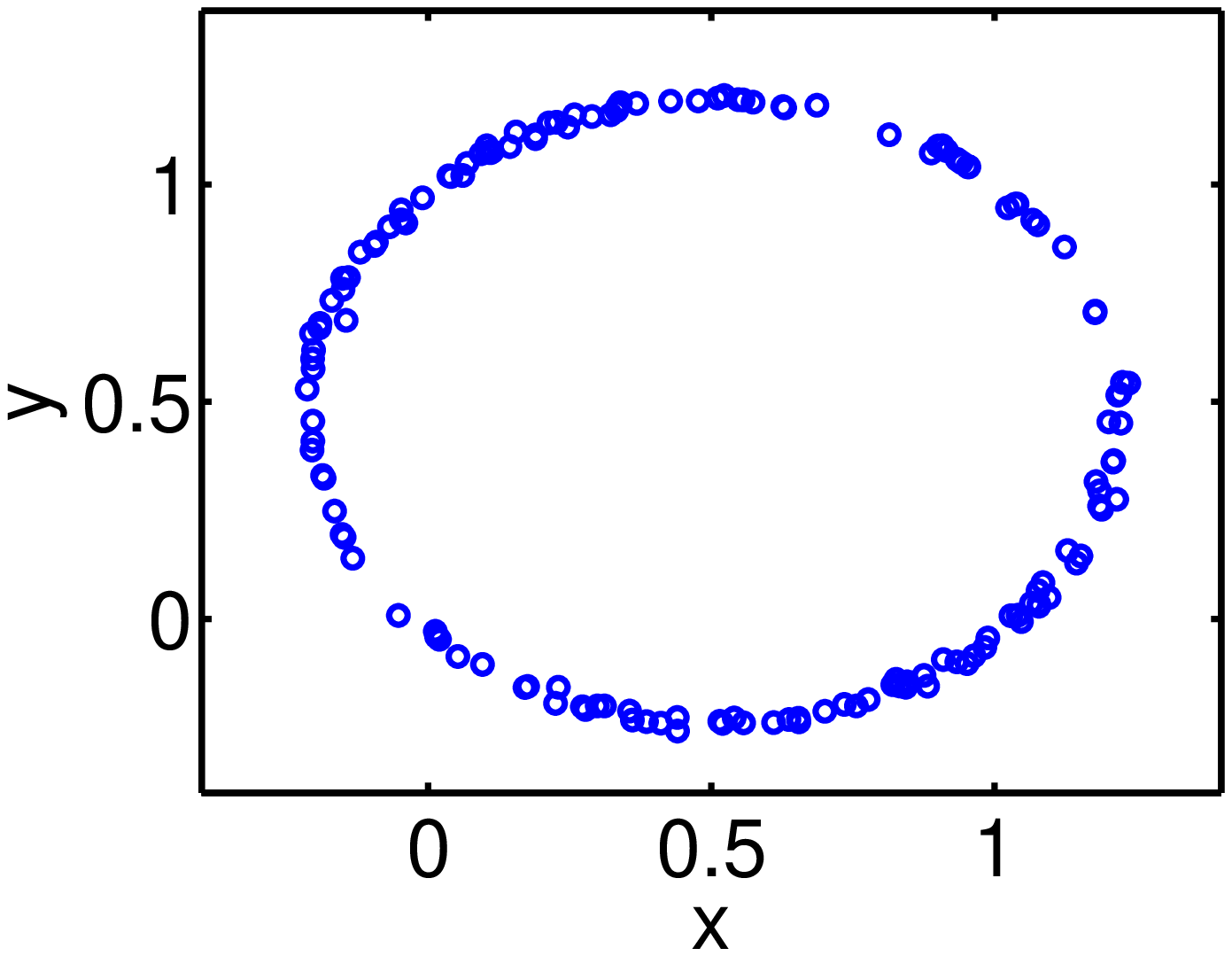}\label{fig1b}}
\caption{Two stable attractors for the swarm dynamics. Here $a=2$,
  $\mu_\tau=2$, and $\sigma_\tau=0.15$. The number of particles is set to be
  $N=150$. The final state shown for both simulations is $t=300$. Panel (a)
  depicts the rotating state at three snapshots at times $t=$297.6, 298.8, 300,
  in red, green and blue, respectively. Panel (b) depicts the ring state.}\label{fig1}
\end{figure}

Figure \ref{fig1} shows the two final particle distributions after transients
in a simulation with an initial state of $N=150$ randomly placed particles,
and where $\mu_\tau=2$ and
$\sigma_\tau=0.15$. In this case, depending upon the random selection of
delays, either stable solution (ring or rotating) is possible. To understand
the effects of increasing the standard deviation of the random delays, we use
a Monte Carlo method. At 100 different values of $\sigma_\tau$ in the range
$0\leq \sigma_\tau\leq0.5$, we generate random time delays from the
distribution in Eq. \eqref{rho_tau}, we then simulate the system starting from the same
initial condition and we determine what state is acquired by the swarm in the long-time limit. To determine this, we first measure the time-averaged distance of particle $j$ to
the center of mass over the interval $(t_1, \ t_2)$:
\begin{equation}\label{time_avg_j}
\langle\delta\mathbf{r}_j\rangle_{(t_1,t_2)} =
  \frac{1}{t_2-t_1}\int_{t_1}^{t_2}|\delta \mathbf{r}_j(t)|dt
\end{equation}
where the size of the interval $(t_1, \ t_2)$ is long enough to include several
periods of oscillation. The ensemble average of Eq. \eqref{time_avg_j} is then
\begin{equation}
\langle\delta\mathbf{r}\rangle_{(t_1,t_2)} = \frac{1}{N}\sum_{j=1}^{N}\langle\delta\mathbf{r}_j\rangle_{(t_1,t_2)}.
\end{equation}
A value $\langle\delta\mathbf{r}\rangle_{(t_1,t_2)}\sim 1/\sqrt{a}$ will
indicate\footnote{When the delays are uniform, the ring state has a radius of
  $1/\sqrt{a}$ \cite{MierTRO11}.} that the system has converged to the ring state, while
 $\langle\delta\mathbf{r}\rangle_{(t_1,t_2)}\ll 1/\sqrt{a}$ shows that the rotating
 state has been adopted instead\footnote{This is true for the range of values of
   $\sigma_\tau$ considered here.}.

\begin{figure}[t]
\centering
\subfigure[]{\includegraphics*[scale=.26]{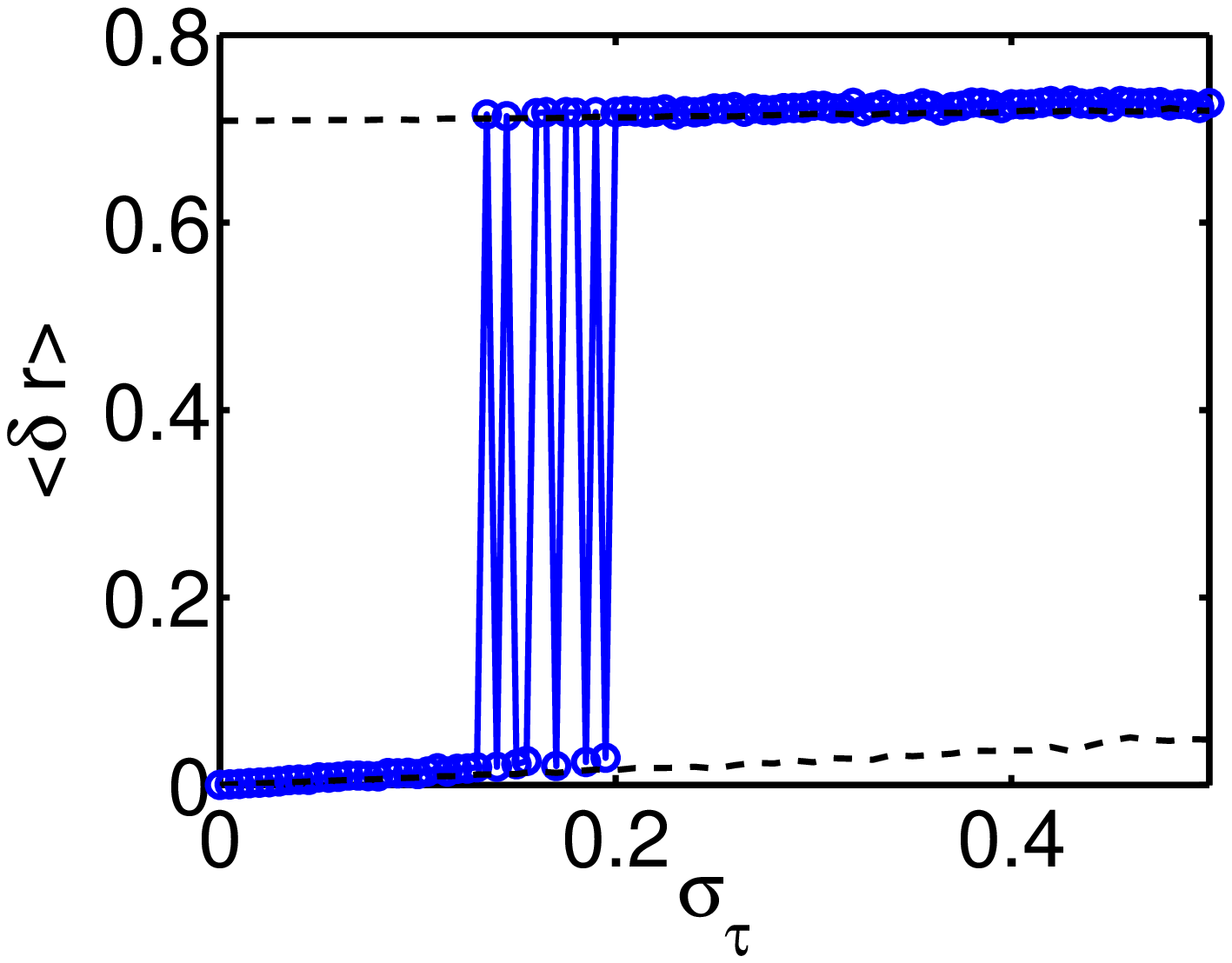}\label{fig2a}}
\subfigure[]{\includegraphics*[scale=.26]{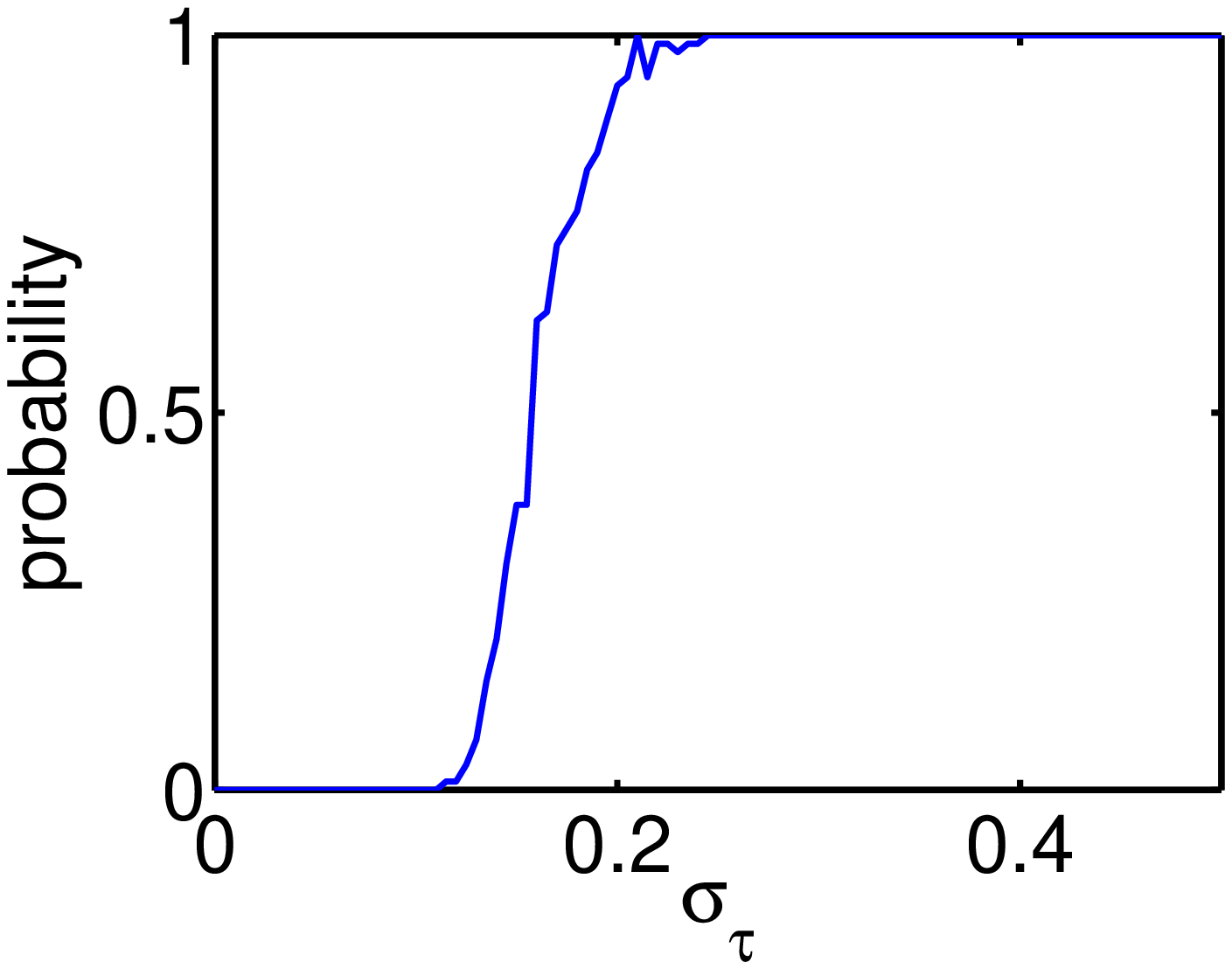}\label{fig2b}}
\caption{As $\sigma_\tau$ is increased, we see a bifurcation from the stable
  rotating state. Panel (a) captures the transition from the rotating
 state to the ring state as the standard deviation of the random delay increases.
 Panel (b) shows the probability of converging to the ring state for a given
 $\sigma_\tau$ of the delays. These results were compiled using a Monte Carlo simulation
with 100 random distributions of delays for 100 uniformly-spaced values of $\sigma_\tau$ and for $N=50$
particles. See accompanying online movie and Appendix to see the agents converge to each
stable pattern.}\label{fig2}
\end{figure}

Figure \ref{fig2a} demonstrates the effect of increasing $\sigma_{\tau}$ on
the final state. The blue circles show that for
$\sigma_\tau$ small, this initial condition converges to the rotating
state. However, for $\sigma_\tau \gtrsim 0.2$ the same initial collection of
particles will converge to the ring state with high probability. In between,
there is a transition region where both states are commonly observed;
the state that occurs depends on the random choice of time delays. The black dashed lines of \ref{fig2a} show two simulations, one which starts near the rotating state
(the lower curve), and one which starts near the ring state (the upper curve)
as $\sigma_\tau$ is increased. These curves demonstrate the stability of these
steady states, and the effect of random delays near these states.

Figure \ref{fig2b} shows the conditional probability of ending up in the ring
state as a
function of $\sigma_\tau$. As expected, for this choice of initial conditions,
for $\sigma_\tau$ small enough, there is zero probability of leaving the rotating state; however, as $\sigma_\tau$ is increased, the probability
increases to one.

The results of these numerical studies strongly suggest that even though there
is bi-stability between the ring and the rotating states, the size of their
respective basins of attraction is changing dramatically as the standard
deviation $\sigma_\tau$ increases.

\addtolength{\textheight}{-7.9cm}

\section{Discussion}

In this paper we studied the dynamics of a self-propelling swarm with
time-delayed inter-agent attraction. In contrast to the previously considered
case of uniform time delay across agents, we considered the situation in which
the time delay between every pair of agents is drawn randomly from a
distribution $\rho_\tau$. 

Using a mean-field model of the swarm, we showed how the two parameter
bifurcation plane of coupling strength and mean time delay changes with
respect to the case in which all time delays are equal. The full implications
of these bifurcation results are the subject of our ongoing work. In
particular, it is unclear what the stable solutions are. Nevertheless, the
dramatic changes seen in the two parameter bifurcation plane as the standard
deviation $\sigma_\tau$ increases suggest that the basins of attraction of
each attractor undergo big changes as well.

Our numerical experiments show that the swarm displays bi-stable behavior
between the ring and rotating states, at the parameters
considered. Interestingly, however, our work suggests that the basin of
attraction of the ring state greatly expands as the distribution of time-delays
$\rho_\tau$ widens. Thus, in a sense, widening the distribution of time-delays
stabilizes the stationary state of the swarm center of mass.

Even though in our model the attractive force among agents is linear, we
believe this work is useful since it represents a first approximation for
other, more general forms of attractive interaction. {Here, we have limited
  our focus to the case where the delays between agents are symmetric and
  constant. However, one important generalization of this system involves
  incorporating time dependent delays, including those which vary as a
  function of the distance between the two agents. This particular refinement
  of our model is the subject of ongoing work and beyond the scope of the
  current paper.} 

{Finally, although we did not consider repulsion between agents,
preliminary research leads us to believe that the patterns observed in this
investigation persist when the characteristic repulsion strength between
robots is small compared to global attraction parameters.} For these reasons,
our results indicate how to exploit time-delayed actuation when designing swarm robotic systems with desired tasks and functionalities.

\section{Appendix-Video Description}

The purpose of this research is to investigate the effects of randomized
communication delay
on emerging patterns in swarming dynamics. This short video captures the
transition between two different stable patterns for a swarm as a function of
the standard distribution of the delays. 

The two coordinate axes in the video show a scatter plot of the positions of
the particles animated in time. The initial positions are identically randomly
distributed  particles in the unit box. The temporal state of the
swarm is updated in time using a numerical scheme called Heun's Method, and a
snapshot is captured at every discrete time interval. Here, the left
coordinate axis uses a standard deviation of the delays $\sigma_\tau=0.1$ while
the right axis uses a standard deviation of $\sigma_\tau=0.3$. The mean delay
for each simulation is set to be $\mu_\tau=2$, and the number of particles for both
is $N=50$. The vectors at each particle give the velocity associated
with that particle. The two
simulations are run side by side to demonstrate the dynamics involved in
converging to the ``rotating'' final state on the left, and the ``ring'' final
state on the right. The video demonstrates the dynamics over the time interval
from $t=0$ to $t=45$, and so includes transients.

\section{ACKNOWLEDGMENTS}

The authors gratefully acknowledge the Office of Naval Research for their
support. LMR and IBS are supported by Award Number R01GM090204 from the
National Institute Of General Medical Sciences. The content is solely the
responsibility of the authors and does not necessarily represent the official
views of the National Institute Of General Medical Sciences or the National
Institutes of Health.  E.F. is
supported by the Naval Research Laboratory (Award No. N0017310-2-C007).

\bibliographystyle{ieeetr}

\end{document}